\documentclass{ws-procs9x6-cpt16}

\def\etal {{\it et al.}}

\begin{document}

\title{Search for Lorentz Violation in km$^3$-Scale Neutrino Telescopes}

\author{C.A.\ Arg\"uelles,$^1$ G.H.\ Collin,$^1$ 
J.M.\ Conrad,$^1$ T.\ Katori,$^2$ and A.\ Kheirandish$^3$}

\address{
$^{1}$Physics Department, Massachusetts Institute of Technology\\
Cambridge, MA 02139, USA}

\address{
$^{2}$School of Physics and Astronomy, Queen Mary University of London\\
London E1 4NS, UK}

\address{
$^{3}$Department of Physics and Wisconsin IceCube Particle Astrophysics Center\\
University of Wisconsin, Madison, WI 53706, USA}

\begin{abstract}
Kilometer$^3$-scale neutrino detectors such as IceCube, ANTARES, and the proposed Km3Net neutrino observatory in the Mediterranean have measured, and will continue to characterize, the atmospheric neutrino spectrum above 1 TeV. Such precise measurements enable us to probe new neutrino physics, in particular, those that arise from Lorentz violation. In this paper, we first relate the effective new physics hamiltonian terms with the Lorentz violating literature. Second, we calculate the oscillation probability formulas for the two-level $\nu_\mu-\nu_\tau$ sector. Finally, we comment on some of the challenges and outlook for this analysis.
\end{abstract}

\bodymatter

%\section{Introduction}
\phantom{}\vskip10pt\noindent
Neutrino oscillations arise from the non-alignment of the propagation and the weak-interaction hamiltonian eigenstates. In vacuum, the former basis is given by the neutrino mass eigenstates and scales as $1/E$. This scaling implies that precise observation of the properties of high energy neutrinos should explore small perturbations due to new physics. Such cases have been studied in the context of astrophysical neutrinos.\cite{Arguelles:2015dca,Katori:2016eni,Bustamante:2015waa,Diaz:2013wia} Furthermore, the high energy component of the atmospheric neutrino spectrum --- before the onset of the astrophysical component --- is well understood and can also be used to look for small perturbations in the standard hamiltonian. In this work we calculate the oscillation probabilities that affect those high energy atmospheric neutrinos in the presence of Lorentz violation.

%\section{Neutrino oscillations in the presence of Lorentz violation}
One of the most generic ways to introduce new physics in neutrino oscillations is to extend the hamiltonian that drives neutrino oscillations in the following way,
\begin{equation}
  H=H_{\rm std}+\sum_n \left(\frac{E}{\Lambda_n}\right)^n\tilde U_n^\dagger O_n \tilde U_n,
\label{eq:newphysics}
\end{equation}
where the first term is the standard neutrino hamiltonian and the second term has the following components:
$O_n=\mathrm{diag}(O_{n,1},O_{n,2},O_{n,3})$ and $\tilde U_n$ is a unitary matrix. $O_n$ and $\Lambda_n$ set the scale of the new physics and $\tilde U_n$ is the mixing matrix describing the new physics flavor structure.\cite{Arguelles:2015dca,Katori:2016eni} Such new operators are explicitly introduced in the context of Lorentz symmetry violation and, in general, depend on the neutrino four-momentum, $p^\mu$. In the minimal Standard-Model Extension (SME), which restricts itself to dimension-four operators, only two terms are introduced:\cite{Kostelecky:2003cr} those corresponding to $n=0$ and $n=1$. In the SME notation they are, respectively, denoted by $a^\lambda_{\alpha\beta}$ and $c^{\lambda\sigma}_{\alpha\beta}$; where $\alpha,\beta = e,\mu,\tau$ are indices that dictate the flavor structure and $\lambda,\sigma$ are Lorentz indices.
These terms can be explicitly written in the following way
\begin{equation}
H=
H_{\rm std}
+
\frac{p_\lambda}{E}
\left( \begin{array}{ccc}
a^\lambda_{ee} & a^\lambda_{e\mu} & a^\lambda_{e\tau} \\
a^{\lambda^*}_{e\mu} & a^\lambda_{\mu\mu} & a^\lambda_{\mu\tau} \\
a^{\lambda^*}_{\mu\tau} & a^{\lambda^*}_{e\tau} & a^\lambda_{\tau\tau} \end{array} \right) 
- \frac{p_\lambda p_\sigma}{E}
\left( \begin{array}{ccc}
c^{\lambda\sigma}_{ee} & c^{\lambda\sigma}_{e\mu} & c^{\lambda\sigma}_{e\tau} \\
c^{\lambda\sigma^*}_{e\mu} & c^{\lambda\sigma}_{\mu\mu} & c^{\lambda\sigma}_{\mu\tau} \\
c^{\lambda\sigma^*}_{\mu\tau} & c^{\lambda\sigma^*}_{e\tau} & c^{\lambda\sigma}_{\tau\tau} \end{array} \right), 
\label{eq:SMEhamiltonian_full}
\end{equation}
where $p_\lambda = (E,\vec p)$ is the neutrino four-momentum. In this work we will impose the simplifying assumption that  $a^\lambda_{\alpha\beta}$ and $c^{\lambda\sigma}_{\alpha\beta}$ are isotropic tensors, i.e., they only have time components.\cite{Diaz:2011ia} In this scenario, the spatial dependence is neglected and is called isotropic Lorentz violation. Further, as the terms need to be traceless in the Lorentz indices, a factor of 4/3 needs to be included. With these assumptions, Eq.\ \eqref{eq:SMEhamiltonian_full} simplifies to
\begin{equation}
H = H_{\rm std} + a_{\alpha\beta} - \tfrac{4}{3} E c_{\alpha\beta}.
\label{eq:SMEhamiltonian_simple}
\end{equation}
Under this simplification, the relationship between Eqs.\ \eqref{eq:newphysics} and \eqref{eq:SMEhamiltonian_full} is now apparent. In order to make it more comparable we choose to redefine $-4 c_{\alpha\beta}/3 \to c_{\alpha\beta}$ in what follows in this paper.

Having introduced these new terms in the hamiltonian, we must now consider where they are relevant. The standard hamiltonian comprises two pieces: $H_{\rm vac}$ and $H_{\rm matter}$. The first term scales like $\Delta m^2/E$, while the second one is proportional to the matter density. The largest squared-mass-difference makes $H_{\rm vac} \sim 10^{-24}({\rm TeV}/{E})$ GeV.  Within the Earth, $H_{\rm matter}$ has only one relevant component which is $H_{\rm matter}^{ee} \sim 10^{-23}$ GeV. If we then restrict ourselves to scenarios in which we consider only the $\nu_\mu$-$\nu_\tau$ sector, then only the $H_{\rm vac}$ scaling matters. This scenario leads to a back of the envelope SME parameter sensitivity estimation of  $a \sim 10^{-24}$-$10^{-27}$~GeV and $c \sim 10^{-27}$-$10^{-32}$, where the upper and lower sensitivity ranges correspond to neutrino energies of 1 TeV and 1 PeV.

\begin{figure}[b]
\centering
\includegraphics[width=0.49\hsize,clip]{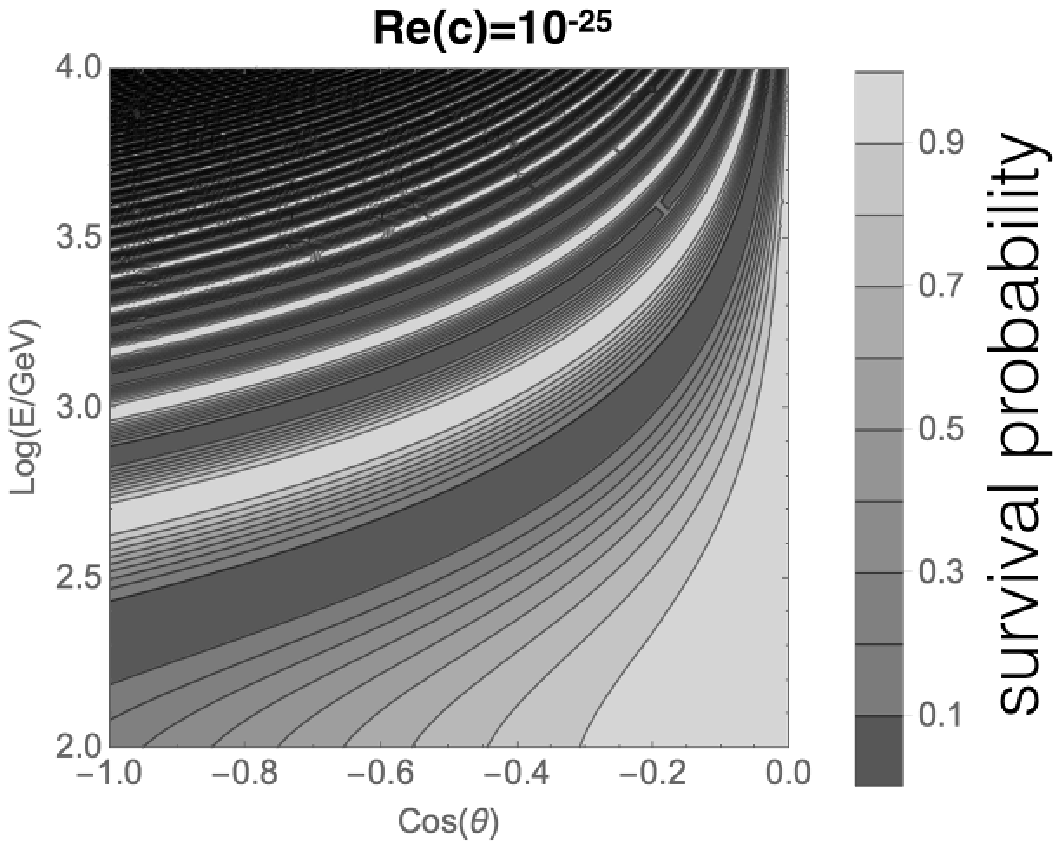}
\includegraphics[width=0.49\hsize,clip]{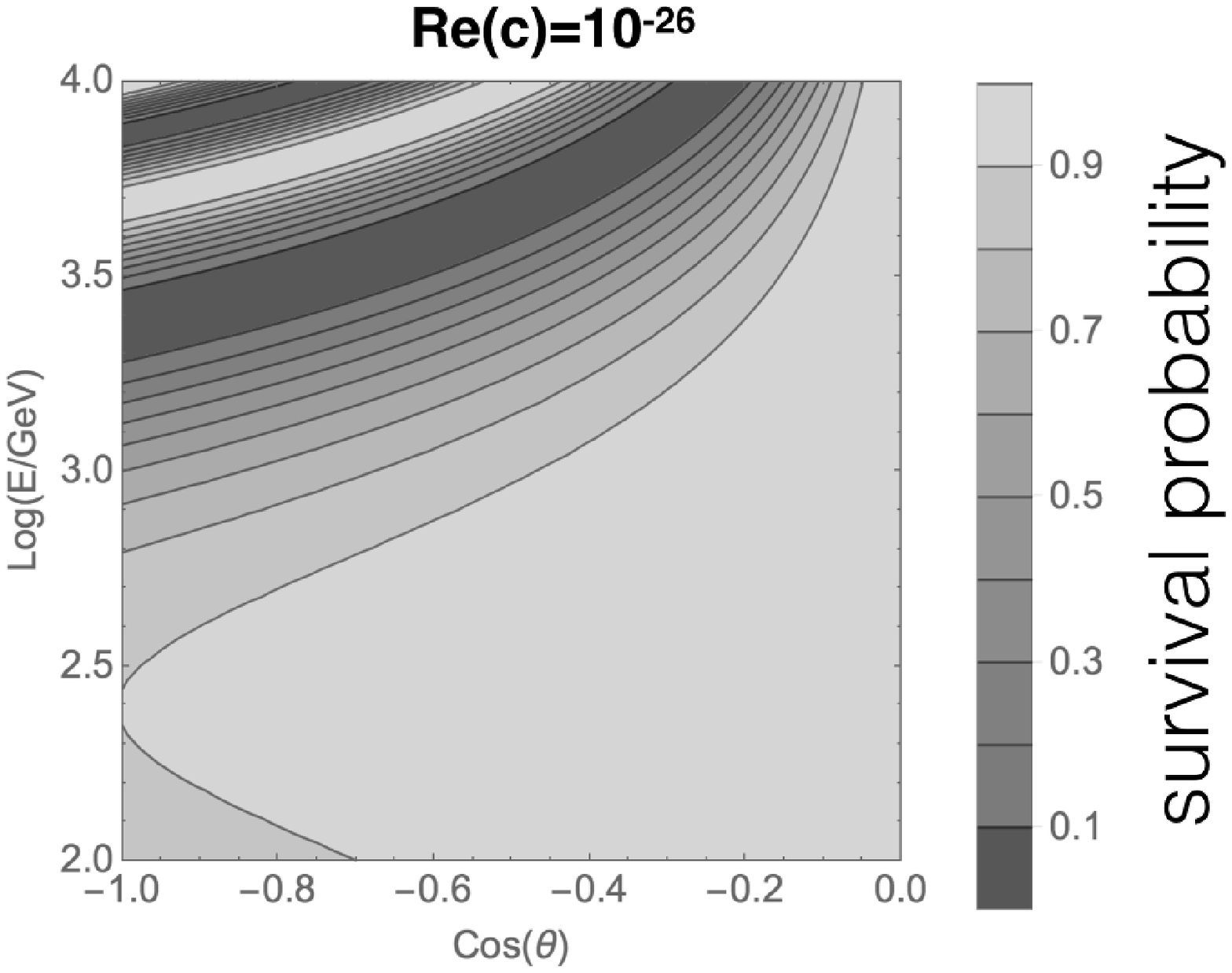}
\caption{\label{fig:oscillations}The left (right) figure shows the $\nu_\mu$ survival probability calculated according to Eq.\ \eqref{eq:osc_prob} for $Re(c_{\mu\mu}) = 10^{-25} (10^{-26})$.}
\end{figure}

Now that we have established the scale of our search,  we proceed to calculate the oscillation probabilities in this regime. Since, in the $\nu_\mu$-$\nu_\tau$ sector, the matter potential drops out of the hamiltonian we can write it in the following way,\cite{GonzalezGarcia:2005xw}
\begin{equation}
H = 
\frac{1}{2E} U^\dagger(\theta) 
\left(
\begin{array}{cc}
0 & 0  \\
0 & \Delta m^2
\end{array}
\right)
U(\theta)
+ E^n
\left(
\begin{array}{cc}
\delta_{\mu\mu} & \delta_{\mu\tau}  \\
\delta^{*}_{\mu\tau} & -\delta_{\mu\mu}
\end{array}
\right),
\label{eq:SMEhamiltonian_two_level}
\end{equation}
where $\delta_{\alpha\beta}$ is the CPT conserving $a$-term, for $n=0$, and the CPT violating $c$-term, for $n=1$, respectively. Then the transition probabilities are\cite{GonzalezGarcia:2005xw}
\begin{equation}
P_{\nu_\mu \to \nu_\mu} = 1 - \sin^22\Theta \sin^2 \left(\frac{\Delta m^2 L}{4E} R\right),
\label{eq:osc_prob}
\end{equation}
where 
\begin{eqnarray}
\sin^22\Theta &=& \frac{1}{R^2}(\sin^22\theta+R_n^2\sin^22\xi+2R_n\sin2\theta \sin2\xi \cos\eta), \\
R &=& \sqrt{1+R_n^2+2R_n(\cos2\theta \cos2\xi+\sin2\theta \sin2\xi \cos\eta)},
\end{eqnarray}
with 
\begin{eqnarray}
R_n &=& \sqrt{\delta_{\mu\mu}^2+{\rm Re}(\delta_{\mu\tau})^2+{\rm Im}(\delta_{\mu\tau})^2}\frac{4E^{n+1}}{\Delta m^2}, \\
\tan \eta &=& \frac{{\rm Im}(\delta_{\mu\tau})}{{\rm Re}(\delta_{\mu\tau})},
\quad 
\tan 2 \xi = \frac{|\delta_{\mu\tau}|}{\delta_{\mu\mu}}
\label{eq:osc_prob_aux_2}.
\end{eqnarray}
In Fig.\ \ref{fig:oscillations} we show the effect on $\nu_\mu$ disappearance for two values of $c$; results with $a$-terms have similar features.

%\section{Outlook and summary}
In summary,
we start from a general parametrization of new neutrino oscillation physics, as introduced, e.g., in Arg\"uelles \etal,\cite{Arguelles:2015dca} and relate the the terms that arise from Lorentz violation. Then, we explicitly calculate the oscillation probabilities in the $\nu_\mu$-$\nu_\tau$ sector. The next step of this work requires modelling the event expectation for experiments that measure the high energy atmospheric neutrino component, such as IceCube and ANTARES. We then need to proceed to include systematic errors associated with the atmospheric neutrino fluxes, as considered in Ref.\ \refcite{TheIceCube:2016oqi}. Then, we will proceed to search for the existence of Lorentz violation in high energy $\nu_\mu$ data.

\section*{Acknowledgments}

C.A.\ and G.C.\ thank the organizers of the CPT'16 conference 
for their hospitality during their stay in Bloomington. 
G.C., C.A., and J.C.\ are supported by NSF grants 1505858 and 1505855.
T.K.\ is supported by the Science and Technology Facilities Council, UK. 
A.K.\ was supported in part by the NSF under grants 0937462 and 1306958 
and by the University of Wisconsin Research Committee.

\end{document}